\documentclass[%
 reprint,
superscriptaddress,
 amsmath,amssymb,
 prl,
]{revtex4-1}
\usepackage{braket}
\usepackage{xcolor}
\usepackage{lipsum}
\usepackage{mathrsfs}
\definecolor{darkblue}{rgb}{0.07, 0.04, 0.56}
\DeclareMathOperator\erf{erf}
\DeclareMathOperator\sinc{sinc}
\DeclareMathOperator\Tr{Tr}
\DeclareMathOperator\Exp{exp}
\usepackage[colorlinks=true,allcolors=blue]{hyperref}
\usepackage{graphicx}
\usepackage{dcolumn}
\usepackage{bm}

\usepackage{amstext}

\begin{document}

\preprint{APS/123-QED}

\title{Dual-Pump Approach to Photon-Pair Generation: Demonstration of Enhanced Characterization and Engineering Capabilities}
\author{Yujie Zhang}
\affiliation{Department of Physics, University of Illinois at Urbana-Champaign, Urbana, Illinois 61801, USA}%
\author{Ryan Spiniolas}
\affiliation{Department of Physics, University of Illinois at Urbana-Champaign, Urbana, Illinois 61801, USA}%
\author{Kai Shinbrough}
\affiliation{Department of Physics, University of Illinois at Urbana-Champaign, Urbana, Illinois 61801, USA}%
\author{Bin Fang}
\altaffiliation[Present address: ]{The Center for Dynamics and Control of Materials: an NSF MRSEC, The University of Texas at Austin, TX 78712, USA}
\affiliation{Department of Physics, University of Illinois at Urbana-Champaign, Urbana, Illinois 61801, USA}%
\author{Offir Cohen}
\affiliation{Department of Physics, University of Illinois at Urbana-Champaign, Urbana, Illinois 61801, USA}%
\affiliation{Frederick Seitz Materials Research Laboratory, University of Illinois at Urbana-Champaign, Urbana, Illinois 61801, USA}
\author{V. O. Lorenz}
\email{vlorenz@illinois.edu}
\affiliation{Department of Physics, University of Illinois at Urbana-Champaign, Urbana, Illinois 61801, USA}%
\date{\today}

\begin{abstract}
We experimentally study the generation of photon pairs via spontaneous four-wave mixing with two distinct laser pulses. We find that the dual-pump technique enables new capabilities: 1) a new characterization methodology to measure noise contributions, source brightness and photon-collection efficiencies directly from raw photon-count measurements; 2) an enhanced ability to generate heralded single photons in a pure quantum state; and 3) the ability to derive upper and lower bounds on heralded-photon quantum state purity from measurements of photon-number statistics even in the presence of noise. Such features are highly valuable in photon-pair sources for quantum applications.
\end{abstract}

\pacs{Valid PACS appear here}
\maketitle

\section{Introduction}

Optical quantum states for quantum applications are commonly realized through photon-pair generation using spontaneous parametric down-conversion (SPDC), in which one pump photon is annihilated and a photon pair is created via the second-order nonlinearity in an interaction medium, and spontaneous four-wave mixing (SFWM), in which two pump photons are annihilated and a pair is created via a $\chi^{(3)}$ nonlinearity. Much research has been conducted on engineering and optimizing these techniques for quantum applications such as quantum computation~\cite{Knill2001}, quantum metrology~\cite{PhysRevLett.96.010401}, and quantum communication~\cite{Gisin2007}. Important figures of merit include the pair-generation probability and the collection efficiency, which determine bit rates and signal-to-noise ratio, and false detections, which degrade the quality of the photon-pair source. Despite the importance of these parameters, in many sources noise contributions prohibit their direct assessment, and to date such photon-pair source characteristics have been calculated by neglecting noise or assuming pump-independent noise contributions~\cite{Smirr:11}, indirect measurements of various source performances~\cite{alibart_photon_2006,Zhou:09,Eckstein2011}, or from fits to multiple measured data points~\cite{Dyer:08,Park2018,Engin:13}. Another figure of merit that is critical for protocols and gates that rely on interference of photons from separate sources is the degree of quantum state purity of the individual photons.

Spontaneous four-wave mixing, unlike the SPDC process, can occur with two spectrally distinct pump fields (see Fig.~\ref{fig:setup}(a)); this additional degree of freedom is beneficial for photon-pair source design, with experimental uses including degenerate photon-pair generation \cite{Fan:05,Medic:10,Silverstone2013,He2014} and avoiding single-pump SFWM background \cite{PhysRevLett.120.233601}. In addition, it has been shown theoretically that dual-pump SFWM leads to improved capabilities in tailoring the inter-correlations of the photon pairs \cite{Mcguinness2007,Monroy:16}, including a proposed method that relies on the group-velocity difference between pump pulses \cite{Fang2013}.

Here we report experimental demonstrations of some key advances in photon-pair generation in general, and dual-pump SFWM in particular. First, we show how the dual-pump scheme enables a simple and direct measurement of the noise contribution to the detection events; this noise consists of background photons from ambient light, photons from additional processes that occur concurrently with photon-pair generation, or false detection events. In turn, measurement of the noise contribution allows a direct quantitative assessment of source performance, including the photon-pair generation rate as well as overall collection and detection efficiencies of the created photons~\cite{alibart_photon_2006}. Second, we show that the group delay between the two pump pulses enables the creation of photon pairs where each of the individual photons is in a highly pure quantum state~\cite{Fang2013}, and third, we derive a new way to determine the lower and upper bounds for the individual photon purity from second-order coherence measurements in the presence of noise -- a method that naturally applies to the dual-pump SFWM where the noise can be directly characterized, but may also find use to estimate other types of photon-pair sources.

This Paper is organized in the following way: in Section~\ref{sec:background} we give an overview of photon-pair generation in dual-pump SFWM, including generation probability and a description of the quantum state. In Section~\ref{sec:generation} we demonstrate and analyze photon-pair production with dual pumps. In Section~\ref{sec:purity} we experimentally confirm the advantage of the dual-pump scheme in generating photon pairs with reduced spectral correlations, both through joint spectral density and single-photon purity measurements. Finally, in Section~\ref{sec:conclusion} we conclude and discuss our results.

\section{Background -- photon-pair state produced in dual-pump SFWM}\label{sec:background}
In the general process of dual-pump SFWM (Fig.~\ref{fig:setup}(a)) two distinct laser pulses -- which we designate as pump 1 and pump 2 with carrier angular frequencies $\omega_{p1}$ and $\omega_{p2}$, respectively -- enter a $\chi^{(3)}$ medium where one photon from each pump pulse is annihilated and signal and idler photons, with carrier angular frequencies $\omega_{s}$ and $\omega_{i}$, respectively, are created as a photon pair (conventionally $\omega_s > \omega_i$). A temporal delay $\tau$ may be applied to pump 1 relative to pump 2. The carrier frequencies  $\omega_s$ and $\omega_i$ are determined by the energy conservation constraint $\omega_{p1}+\omega_{p2}=\omega_{s}+\omega_{i}$ as well as the phasematching conditions that are specific to the $\chi^{(3)}$ medium.
For this work we choose polarization maintaining fiber (PMF) as the nonlinear medium, where the polarization of both pump pulses are aligned with the slow axis of the PMF while the signal and idler photons are generated with polarization along the fast axis. In such a design the phasematching conditions are given by~\cite{Fang2013}: 
\begin{equation}\label{eq:phasematching}
\Delta k=k(\omega_{p1})+k(\omega_{p2})-k(\omega_{s})-k(\omega_{i}) + \Delta n \frac{\omega_{p1} + \omega_{p2}}{c} =0,
\end{equation}
where $k(\omega)=n(\omega)\omega/c$, $n(\omega)$ is the fast-axis (effective) refractive index in the fiber, which we model based on the Sellmeier equation of bulk silica~\cite{Smith2009}, $\Delta n$ is the fiber birefringence, and $c$ is the speed of light in vacuum. Starting with the $\ket{\textrm{vac}}$ state, in which no photons exist in either signal or idler modes, the output state of the dual-pump SFWM process can be evaluated perturbatively as~\cite{Mcguinness2007}:
\begin{equation}
    \ket{\Psi} = \ket{\textrm{vac}} + \kappa \iint d\nu_s d\nu_i F(\nu_s, \nu_i; \tau) \ket{\nu_s, \nu_i} + \mathcal{O}(\kappa^2),
    \label{eq:state}
\end{equation}
where $\ket{\nu_s, \nu_i}$ represent a photon-pair state in which the signal (idler) angular frequency is $\omega_{s(i)} + \nu_{s(i)}$ and $\kappa$ is the interaction coupling constant that depends on the relevant $\chi^{(3)}$ nonlinear susceptibility, the fiber length $L$, and pump powers, but not on the time delay $\tau$ that is applied to pump 1 relative to pump 2 before they are coupled into the fiber. The unnormalized joint spectral amplitude is given by~\cite{Mcguinness2007,Fang2013}: 
\begin{widetext}
\begin{equation}
\begin{split}
&F(\nu_s,\nu_i;\tau)
=\exp{\left[-\frac{(\nu_s+\nu_i)^2}{\sigma_1^2+\sigma_2^2}\right]}\exp{\left[-\left(\frac{T_s\nu_s+T_i\nu_i}{\sigma\tau_p}\right)^2\right]}\\
&\times\left[\erf\left(\frac{\sigma(\tau+\tau_p)}{2}-i\frac{T_s\nu_s+T_i\nu_i}{\sigma\tau_p}\right)-\erf\left(\frac{\sigma\tau}{2}-i\frac{ T_s\nu_s+T_i\nu_i}{\sigma\tau_p}\right)\right].
\label{eqn:JSD}
\end{split}
\end{equation}
\end{widetext}
Here $\sigma_{1(2)}$ denotes the pump 1 (pump 2) spectral bandwidth (half width at $1/e^2$ maximum amplitude); $\sigma=\sigma_1\sigma_2/\sqrt{\sigma_1^2+\sigma_2^2}$; $T_{s(i)}=\tau_{s(i)}+(\sigma_1^2-\sigma_2^2)/(\sigma_1^2+\sigma_2^2)\tau_p/2$, where  $\tau_{s(i)}=L((k'_{p1}+k'_{p2})/2-k'_{s(i)})$ is the group delay difference between the signal (idler) and the average group delay of the pumps acquired during the propagation in the fiber, $k'_{s(i)}=dk/d\omega|_{\omega_{s(i)}}$ is the inverse group velocity of the signal (idler) in the fiber, $\tau_p=L(k'_{p1}-k'_{p2})$ is the group delay between the two pumps acquired during the propagation in the fiber, and $k'_{p1(2)}=dk/d\omega|_{\omega_{p1(2)}} + \Delta n/c$ is the inverse group velocity of pump 1 (2) in the fiber. The probability $p(\tau)$ that a photon pair is generated, $p(\tau)=|\kappa|^2\times \iint d\nu_s d\nu_i|F(\nu_s,\nu_i;\tau)|^2$, is given by
\begin{equation}
p({\tau})=p_{\mathrm{max}}\left[\frac{\erf{\left(\frac{\sigma\tau+\sigma\tau_p}{\sqrt{2}}\right)}-\erf{\left(\frac{\sigma\tau}{\sqrt{2}}\right)}}{\erf{\left(\frac{\sigma\tau_p}{2\sqrt{2}}\right)}-\erf{\left(-\frac{\sigma\tau_p}{2\sqrt{2}}\right)}}\right],
\label{eq:pt}
\end{equation}
where $p_{\mathrm{max}}$ corresponds to the maximum generation probability, which occurs when pump 1 and pump 2 maximally overlap in the middle of the fiber, i.e., $\tau=-\tau_p/2$.

\section{Demonstration and characterization of photon-pair generation}\label{sec:generation}

\begin{figure}[t]
\centering
\includegraphics[width=0.8\linewidth]{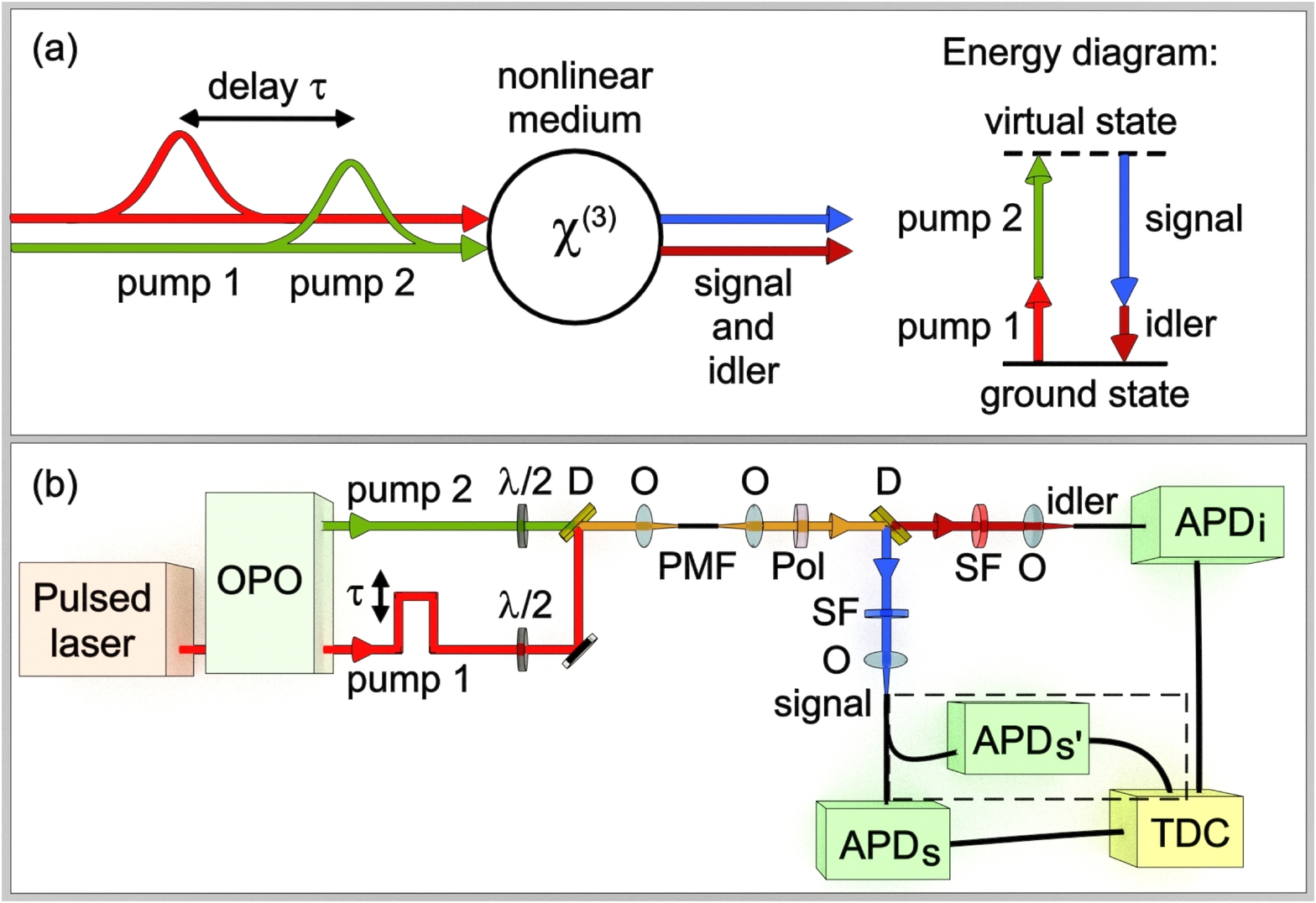}
\caption{(a) Schematic of the dual-pump SFWM process. (b) Experimental setup for dual-pump SFWM. Different combinations of avalanche photo-diodes $\text{APD}_s$, $\text{APD}_{s^\prime}$ and $\text{APD}_i$ are used, as described in the text; the dashed box indicates a fiber beam-splitter and $\text{APD}_{s^\prime}$ are present only for those measurements that require it. OPO: optical parametric oscillator; $\lambda/2$: half-wave plate; O: objective; PMF: polarization-maintaining fiber; Pol: polarizer; D: dichroic mirror; SF: spectral filter; TDC: time-to-digital converter.}
\label{fig:setup}
\end{figure}

We study the statistical properties of photon pairs generated in the dual-pump SFWM scheme using the experimental setup shown in Fig.~\ref{fig:setup}(b). A Ti:sapphire modelocked laser with $80\,$MHz repetition rate, $772\,$nm central wavelength and $8\,$nm full-width-at-half-maximum (FWHM) bandwidth, pumps an optical parametric oscillator (OPO). The residual pump output from the OPO is used as pump 1. Pulses are generated in the OPO at central wavelengths between $530$-$660\,\textrm{nm}$ with typical FWHM bandwidths of $1$-$3\,\textrm{nm}$, corresponding to pump 2. Pump 1 is time-delayed from pump 2 by an amount $\tau$ using an automated translation stage. The two pump paths are combined on a dichroic mirror and are coupled into a 1.6-cm long polarization-maintaining fiber (PMF) (PM630-HP, birefringence $3.5\times10^{-4}$), which serves as the nonlinear medium in which the SFWM process takes place \cite{Smith2009,soller_high-performance_2011,Fang:14}; the pump polarizations are aligned along the slow axis of the fiber. The signal and idler photons are produced with orthogonal polarization along the fast axis. A polarizer, which allows the signal and idler photons through, rejects most of pumps 1 and 2 and reduces noise from spurious interactions in the fiber. The signal and idler photons are separated by a dichroic mirror and each is coupled into a single-mode fiber connected to avalanche photodiode (APD) single-photon detectors (Excelitas SPCM-AQ4C). The detection signals, which are represented by electronic pulses from the APDs, are collected by a time-to-digital converter (TDC) that time-stamps and processes detection events to record the number of single detection events $C_{s}$, $C_{i}$ and $C_{s^\prime}$ at $\text{APD}_{s}$, $\text{APD}_{i}$ and $\text{APD}_{s^\prime}$, respectively; two-fold coincidences $C_{s i}$ ($C_{s s^\prime}$) at $\text{APD}_{s}$ \& $\text{APD}_{i}$ ($\text{APD}_{s}$ \& $\text{APD}_{s^\prime}$); and three-fold coincidences $C_{s s^\prime i}$ at $\text{APD}_{s}$ \& $\text{APD}_{s^\prime}$ \& $\text{APD}_{i}$.

\par

\begin{figure}[t]
\centering
\includegraphics[width=0.8\linewidth]{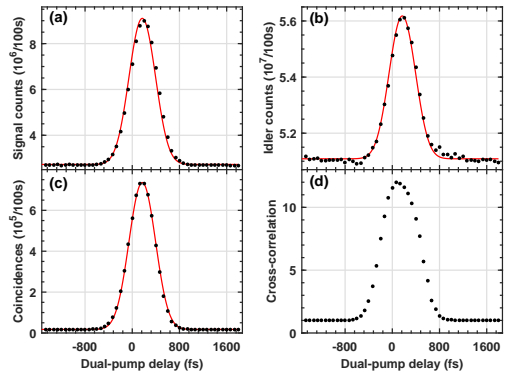} 
\caption{(a-c) Photon detection counts (circles) and multi-curve fits (solid lines) vs.~dual-pump delay for the (a) signal arm ($C_s$), (b) idler arm ($C_i$), and (c) coincidences ($C_{si}$). (d) $g^{(2)}_{si}$ second-order coherence cross-correlation calculated from the counts in (a-c).}
\label{fig:counts}
\end{figure}

With $70\,\textrm{mW}$ average power in pump 1 at $772\,\textrm{nm}$, and pump 2 set at $622\,\textrm{nm}$ with $20\,\textrm{mW}$ power, we record the number of detection events at the signal arm $C_s$, idler arm $C_i$, and coincidence counts $C_{si}$, as a function of time delay $\tau$ between the two pump pulses. The results are presented in Figs.~\ref{fig:counts}(a)-\ref{fig:counts}(c). At a certain delay $\tau=\tau_0$ the counts reach peak values, as expected for dual-pump SFWM, in which the photon-pair generation rate depends on the overlap between the two pump pulses~\cite{Fang2013}. For large $|\tau|$ (where the two pumps do not overlap and thus no dual-pump SFWM occurs) they asymptotically approach non-zero lowest values that amount to background photons and detection events, mainly due to Raman scattering, but also from single-pump SFWM, ambient light and dark counts. Figure \ref{fig:counts}(d) presents the cross-correlation $g^{(2)}_{si}=C_{si}R/C_s C_i$, where $R$ is the number of dual-pump pulse pairs over which the counts are taken (which is the laser repetition rate times the measurement duration). In order to ensure that the counts are synchronized with the laser pulses, the unconditional $C_s$ and $C_i$ counts in Figs.~\ref{fig:counts}(a)-\ref{fig:counts}(b) are gated at the laser repetition rate divided down to $8\,\textrm{MHz}$ (due to bandwidth limitations of the TDC), and have been multiplied by $10$ to reflect counts at the laser repetition rate; these values are used to calculate the cross-correlation $g^{(2)}_{si}$ in Fig.~\ref{fig:counts}(d). At the peak, $g^{(2)}_{si}=11.98\pm0.02$, indicating
that correlated detection events of signal and idler photons occur. For large $|\tau|$, $g^{(2)}_{si}\rightarrow 1$, as expected from Poisson statistics of counts that originate from uncorrelated noise, confirming that signal-idler pairs indeed originate from dual-pump SFWM. We also measure the conditional auto-correlation of signal photons upon idler photon detection, $g_{s s^\prime|i}^{(2)}=(C_{s s^\prime i} C_i/C_{si} C_{s^\prime i})$ (all counts are taken at $\tau=\tau_0$), yielding $g_{s s^\prime |i}^{(2)} = 0.017 \pm 0.002$, which indicates a low probability of multi-photon emission in one arm upon photon detection in the other arm.
\par

Generally, the counts are given by:
\begin{widetext}
\begin{subequations}\label{eq:counts}
    \begin{align}
        C_s(\tau) =&\; N_s + \eta_s p(\tau)R,\\
        C_i(\tau) =&\; N_i + \eta_i p(\tau)R,\\
            C_{si}(\tau) =&\; \frac{N_s N_i}{R} + \left(1-\eta_s\right)p(\tau)N_i  + \left(1-\eta_i\right)p(\tau)N_s + \eta_s\eta_i p(\tau) R,\label{eq:coinc_counts}
    \end{align}
\end{subequations}
\end{widetext}
where $N_s$ ($N_i$) and $\eta_s$ ($\eta_i$) are the noise counts and the detection efficiency (accounting for both collection and detector efficiencies) of the signal (idler) photons generated in the dual-pump SFWM, respectively. In experiment, data is collected at various positions of the delay stage in Fig.~\ref{fig:setup}(b). When the stage is in its central position, pump 1 is delayed relative to pump 2 by an unknown delay $\tau_c$; the position of the stage is translated to create relative temporal delays  $\tau_\mathrm{exp}$ of pump 1. Fitting curves to the data in Figs.~\ref{fig:counts}(a)-\ref{fig:counts}(c) are generated by substituting Eq.~(\ref{eq:pt}) into Eqs.~(\ref{eq:counts}) with $\tau=\tau_\mathrm{exp}-\tau_c$, and simultaneously fitting the three curves to the data, with $N_{s}$, $N_i$, $\eta_{s}$, $\eta_i$, $p_{\mathrm{max}}$, $\sigma$, $\tau_p$ and $\tau_c$ as common fitting parameters to all curves.
\par
 The fitting result gives a maximal photon-pair generation probability per dual-pump pulse of $p_{\mathrm{max}} = (6.0\pm 0.2) \times 10^{-3}$, and collection efficiencies of $\eta_s = 13.4 \pm 0.5\%$ and $\eta_i = 10.7 \pm 0.1\%$. The good agreement between model and experiment supports our approach, but we note that modeling is not required to determine the noise contributions, which can be obtained directly from a single measurement at large $|\tau|$ (where $p(\tau)\rightarrow 0$), or three measurements of counts collected once when only pump~1 is present (no pump~2), once when only pump~2 is present (no pump~1) and once when both are blocked. Generally, quantifying noise enables one to gain information about source performance~\cite{alibart_photon_2006}; here, using Eqs.~(\ref{eq:counts}), knowledge of the noise enables us to extract the source performance from raw counts.
\section{Effect of pump detuning on photon-pair state and single-photon state purity}\label{sec:purity}
Dual-pump SFWM also provides enhanced capabilities in generating photon-pair quantum states with engineered properties \cite{Mcguinness2007, Fang2013}. Generally, the spectral quantum state of a photon pair can be expressed as $\ket{\Phi}=\iint d\nu_s d\nu_if(\nu_s,\nu_i)\ket{\nu_s,\nu_i}$, where $f(\nu_s,\nu_i)$ is the normalized joint spectral amplitude (JSA).  The quantum state of the signal (idler) is then given by the density matrix $\rho_{s(i)}=\mathrm{Tr}_{i(s)}(\ket{\Phi}\bra{{\Phi}})$, where $\mathrm{Tr}_{i(s)}$ represents the partial trace over the idler (signal) degrees of freedom. The purity of the signal and idler photons $P=\mathrm{Tr}(\rho_s^2)=\mathrm{Tr}(\rho_i^2)$ amounts to the degree to which they are in pure quantum states rather than mixed states, and is a critical figure of merit~\cite{Meyer-Scott_2017} in quantum protocols that rely on two-photon interference. Many efforts are being put into engineering the properties of photon pairs \cite{PhysRevLett.115.193601,Ansari:18,Lu:18,Branczyk2011,Graffitti2017}. In particular, one of the most useful states is the factorable state, where the JSA can be written as independent wavefunctions of the signal ($f_s(\nu_s)$) and idler ($f_i(\nu_i)$) photons; that is, $f(\nu_s, \nu_i)=f_s(\nu_s)f_i(\nu_i)$, leading to pure ($P=1$) quantum states of signal and idler photons. Conversely, when the two photons are spectrally entangled ($f(\nu_s, \nu_i)$ is not factorable), $P<1$ and the individual photons are in a mixed state.

\begin{figure}[t]
\centering
\includegraphics[width=0.8\linewidth]{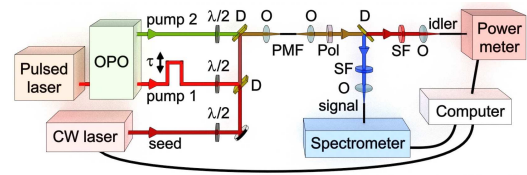}
\caption{Experimental setup for stimulated-emission-based measurement of the joint spectral density. CW: continuous-wave.}
\label{fig:set_setup}
\end{figure}

\subsection{Measurements of the joint spectral density}
\begin{figure*}[t]
 \center
  \includegraphics[width=\textwidth]{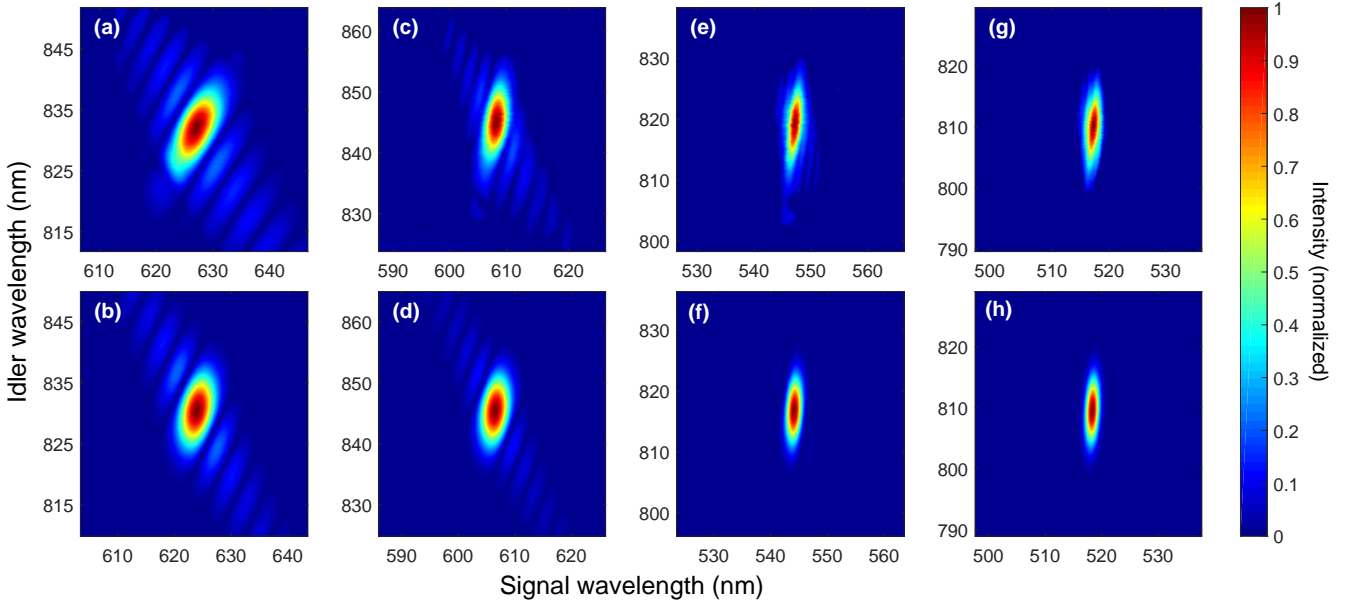}
  \caption{ Experimental (top row) and theoretical (bottom row) joint spectral densities (JSDs) for various detunings. Experimental data is measured via stimulated emission. (a,b) degenerate pump at $715\,$nm, (c,d) dual pump at $772\,$nm and $652\,$nm, (e,f) dual pump at $772\,$nm and $565\,$nm, and (g,h) dual pump at $772\,$nm and $534\,$nm. Going from left to right, corresponding to increasing detuning between the two pumps, the sidelobes' intensity weakens and the JSD of the signal and idler photons becomes less correlated.}
  \label{fig:jsi}
\end{figure*}

To characterize the JSA properties we measure the joint spectral density (JSD), $|f(\nu_s, \nu_i)|^{2}$, using stimulated four-wave mixing,  as proposed in \cite{Liscidini2013} and demonstrated in \cite{Ang2014,Fang:16} (experimental setup shown in Fig.~\ref{fig:set_setup}). This is performed by adding a tunable Ti:sapphire continuous-wave laser that co-propagates with the pumps in the PMF and seeds the idler beam to stimulate the creation of signal-idler pairs. Spectra are collected for each seed wavelength to generate the JSD. In the degenerate (single) pump case the JSA is given by~\cite{Mcguinness2007}:
\begin{equation}
f_{\textrm{degen}}(\nu_s, \nu_i) = \mathcal{N}\exp{\left[-\frac{\left(\nu_s+\nu_i\right)^2}{\sigma_1^2 + \sigma_2^2}\right]} \sinc{(\tau_s \nu_s + \tau_i \nu_i)},
\end{equation}
where $\mathcal{N}$ is a normalization factor. The measured JSD for this case is shown in Fig.~\ref{fig:jsi}(a). It is evident that in addition to the main peak, there are sidelobes; these are due to the wings of the sinc function in $f_{\textrm{degen}}(\nu_s, \nu_i)$, and originate from the sudden onset and ending of the nonlinear interaction when the pump pulse enters and exits the fiber. The fiber length $L\approx 1.6~\textrm{cm}$ in our experiments is chosen based on the model such that the purity of the photons is the highest it can be in the single-pump configuration, reaching a value of $\sim83\%$. While this purity is high considering that no narrow spectral filtering is applied, the sidelobes seen in Fig.~\ref{fig:jsi}(a) -- which constitute strong correlations between the signal and idler photons -- limit the ability to achieve a factorable state~\cite{soller_high-performance_2011}.

In the dual-pump SFWM experiments, the time delay between pump 1 and 2 is set to be $\tau = \tau_0 = -\tau_p/2$ such that the two pumps maximally overlap at the center of the fiber and thus photon-pair production probability is highest. In this case Eq.~(\ref{eqn:JSD}) yields the JSA 
\begin{equation}\label{eq:f}
\small
\begin{split}
&f(\nu_s,\nu_i)=\mathcal{N}\exp{\left[-\frac{(\nu_s+\nu_i)^2}{\sigma_1^2+\sigma_2^2}\right]}\exp\left[-\left(\frac{T_s\nu_s+T_i\nu_i}{\sigma\tau_p}\right)^2\right]\\
&\;\times\left[\erf\left(\frac{\sigma\tau_p}{4}-i\frac{T_s\nu_s+T_i\nu_i}{\sigma\tau_p}\right)-\erf\left(-\frac{\sigma\tau_{p}}{4}-i\frac{ T_s\nu_s+T_i\nu_i}{\sigma\tau_p}\right)\right].
\end{split}
\end{equation}
If the temporal walk-off between the pumps is large enough such that they completely sweep across each other within the medium, i.e.~$\sigma \tau_p \gg 1$, the SFWM interaction strength, which is proportional to the overlap between the pumps, varies along the fiber and the JSA becomes~\cite{Fang2013}:$ 
f_{\sigma\tau_p\gg1}(\nu_s,\nu_i)=\mathcal{N}\exp{[-{(\nu_s+\nu_i)^2}/{(\sigma_1^2+\sigma_2^2})]} \exp{[-((T_s\nu_s+T_i\nu_s)/\sigma\tau_p)^2]}$.

Under such conditions the JSA is expressed as the product of two Gaussian functions, which is the ideal expression for obtaining a factorable state~\cite{Queseda_2018} in general and possesses no sidelobes in particular. It becomes more factorable as the quantity $\mathscr{C} = \left(\sigma_1^2 + \sigma_2^2 \right)T_s T_i + \left(\sigma \tau_p \right)^2$
gets smaller, and, in principle, can become completely factorable when $\mathscr{C}=0$. We note that while the condition $\sigma \tau_p \gg 1$ can be relatively easily satisfied by using a long medium, the value of $\mathscr{C}$ depends strongly on the dispersion characteristics of the medium. In PMF, $\sigma \tau_p$ increases with detuning, while $\mathscr{C}$ decreases. Thence, we expect that increasing the detuning between the pumps would result in less correlations in the JSA~\cite{Fang2013}.

With 20~mW average power each in pumps 1 and 2 and 30~mW average power in the seed beam, we obtain the experimental JSDs for various detunings $\Delta=\lambda_1 - \lambda_2$ (where $\lambda_{1(2)}$ is the central wavelength of pump 1 (2)) shown in Fig.~\ref{fig:jsi}. As can be seen, with increased detuning the sidelobes' intensity weakens and the JSD becomes less correlated. Also shown in Fig.~\ref{fig:jsi} (bottom row) are the corresponding calculated JSDs based on the model; the fidelity between measured ($|f_{\mathrm{meas}}(\nu_s,\nu_i)|^2$) and theoretical $|f_{\mathrm{theory}}(\nu_s,\nu_i)|^2$ JSDs~\cite{Smith2009} $\iint d\nu_s d\nu_i \sqrt{|f_{\mathrm{theory}}(\nu_s,\nu_i)|^2|f_{\mathrm{meas}}(\nu_s,\nu_i)|^2}$ is $>95\%$ for all measurements. These results support the models and the feasibility of the dual-pump approach to generating heralded single photons in pure wavepackets.

\subsection{Purity measurements through autocorrelation}
While the JSDs provide useful information about the photon-pair inter-correlations, they do not include details about the joint phase, and thus bear limited information about the individual photon purity. Photon-number statistics provide the purity~\cite{mauerer_how_2009} directly via the unconditional auto-correlation function~\cite{spring_-chip_2013} measured with the setup in Fig.~\ref{fig:setup}(b), $g_{\mathrm{s s^\prime}}^{(2)} |_{\tau}= C_{s s^\prime}(\tau) R/C_s(\tau) C_{s^\prime}(\tau) = 1+P_{\mathrm{meas}}(\tau)$, where $P_{\mathrm{meas}}$ is the measured quantum-state purity. However, this kind of measurement is highly susceptible to noise contributions, which affect the detection statistics, resulting in an inaccurate deduced purity. Here again we find that the dual-pump scheme provides an advantage for quantifying the properties of the source. We derive upper and lower bounds for the purity of the signal photons $P$ in the presence of noise. We assume two different types of noise: 1) Noise that is generated by the interaction of either pump and creates spurious photons at the signal arm together with an additional boson -- this boson could be another photon (e.g., through single-pump SFWM), or a collective excitation in the medium (such as a phonon). We call this type of noise spurious noise. 2) Noise that occurs at the detector, such as dark counts or ambient light. We refer to this kind of noise as detection noise, with associated purity expressed through the auto-correlation function when both pumps are blocked, $P_{\textrm{det}} = (D_{s s^\prime} R/D_{s} D_{s^\prime} - 1)$, where $D_{s(s^\prime)}$ is the detection-noise counts collected at $\text{APD}_{s(s^\prime)}$ and $D_{s s^\prime}$ is the number of coincidences between $\text{APD}_{s}$ \& $\text{APD}_{s^\prime}$, measured with blocked pumps. Given the raw purity $P_\mathrm{raw} = P_{\mathrm{meas}}(\tau_0)$ (since we are interested in measuring the purity when maximal photon-pair generation occurs), we can find bounds for the true purity of the signal photon produced through the dual-pump SFWM (see Appendix):
\begin{equation}\label{eq:purity_ineq}
\begin{split}
P \le &\;\frac{P_{\mathrm{raw}} - t^2 P_{\mathrm{noise}}}{r^2},\\
P \ge &\;\frac{P_{\mathrm{raw}} - t^2 P_{\mathrm{noise}}}{r^2} - \frac{2t}{r^2}\sqrt{P_{\mathrm{raw}}\left(P_\mathrm{noise} - u^2 P_{\mathrm{det}} \right)},
\end{split}
\end{equation}
where $P_{\mathrm{noise}}=P_{\mathrm{meas}}(\infty)$ is an effective purity associated with the total noise and is measured through the auto-correlation function at large $|\tau|$; $r = \sqrt{(1-t_s)(1-t_{s^\prime})}$ and $t = \sqrt{t_s t_{s^\prime}}$ are the geometric averages of the ratios of signal- and noise-detections to the total counts, respectively, with $t_{s(s^\prime)}=C_{s(s^\prime)}(\infty)/C_{s(s^\prime)}(\tau_0)$; and $u = \sqrt{u_s u_{s^\prime}}$ is the geometric average of the ratio of detection-noise to total noise, where $u_{s(s^\prime)}=D_{s(s^\prime)}/C_{s(s^\prime)}(\infty)$. If $P_{\mathrm{noise}}=0$ then $P_{\mathrm{det}}=0$ necessarily; in such a case, or when $t=0$ (no noise), the two bounds merge and the equality holds in Eqs.~(\ref{eq:purity_ineq}). 

\begin{table*}[t]
  \begin{center}
      \caption{Measured raw purity $P_{\mathrm{raw}}$, measured noise purity $P_{\mathrm{noise}}$, ratio $r$ (in our experiments $t \cong 1-r$) of SFWM signal photon counts to total (SFWM + noise) counts, corrected SFWM signal photon purity $P$, and theoretical quantum state purity $P_{\mathrm{theory}}$ for various spectral detunings $\Delta$ between pumps 1 and 2. The statistical errors in the $r$ values are $<10^{-3}$.}  \label{table:purity}
    \begin{tabular}{cccccc}
    \hline
    $\Delta$ (nm) & $P_{\mathrm{raw}}$ (\%)& $P_{\mathrm{noise}}$ (\%) & $r$ & $P$ (\%) & $P_{\mathrm{theory}}$ (\%)  \\
    \hline
    0 & $58.7 \pm 0.6$&-- &  -- &  -- & 82.6 \\
    120 & $38.2 \pm 0.7$&$1\pm2.6$  & 0.66 & $88.2\pm1.7$ &  88.4 \\
    150 & $34.4 \pm 0.8$ &$-0.3\pm1.4$ & 0.62 & $90.7\pm2.2$  &  92.1\\
    187  & $24.5 \pm 0.4$&$0.7\pm2.4$ & 0.50 & $97.4\pm1.7$ & 95.3\\
    \hline
    \end{tabular}
  \end{center}
  \vspace{-0.1in}
\end{table*}
\par

We measure $P_{\text{raw}}$, $P_{\text{noise}}$, $r$ and $t$ for various pump detunings. We find that $P_{\mathrm{noise}}\sim 0$ for all measurements; we thus assume that it is zero and Eqs.~(\ref{eq:purity_ineq}) turn into the equality $P={P_{\mathrm{raw}}}/{r^2}$. The results of these measurements are summarized in Tab.~\ref{table:purity}, together with a comparison to the theory \cite{Fang2013}. The good agreement between $P$ and the model, and the trend of improving purity with detuning, is yet another confirmation for the dual-pump approach as a superior technique to generate signal and idler pairs with each photon in a pure quantum state. The fact that we cannot use the above procedure to find $r$, $t$ and $P$ for the measurements at $\Delta=0$ (single pump centered at $715\,\textrm{nm}$) emphasizes the advantage of the dual-pump scheme, where one can deduce the quantum state purity of the photons that are truly produced in pairs.

\section{Conclusion}\label{sec:conclusion}

In conclusion, we experimentally investigate the generation of photon pairs through SFWM using two spectrally distinct laser pulses. We devise a new technique that utilizes the dual-pump nature for characterizing the performance of the photon-pair source in terms of generation probability, photon-collection efficiency, noise levels and quantum state purity of the individual photons. As examples of potential applications of these capabilities, one can differentiate between degradation of the source medium, changes in the efficiency of collection and variations in ambient light; by scanning time delay, one can also characterize changes in pulse duration and modify the location of maximal pump overlap in the medium to avoid localized defects. Such tools may be especially useful in quantum applications where characterization of source performance and troubleshooting needs to take place periodically and remotely, especially in cases where the source needs to be placed in hard-to-access locations such as space, or in a network with a vast number of sources. In addition, we show that large spectral detuning between the two pump pulses results in the generation of a highly factorable photon-pair state, with single-photon purities up to $97.4\pm1.7\%$ as determined using dual-pump-enabled noise measurements, far exceeding those attainable with a single pump in the same generation medium.

To perform this first demonstration we choose PMF as the nonlinear medium due to its maturity as an SFWM photon-pair source~\cite{Smith2009,soller_high-performance_2011,Fang:14,Fang:16, Garay-Palmett_2016,Cruz-Delgado_2016}, the straightforwardness of the experimental setup and the simplicity of the model, which has a long track record of matching well with experimental results. It is expected, though, that more sophisticated media will be able to better exploit the dual-pump SFWM and overcome some of the issues found in PMF; for example, it has been proposed that with an adequately engineered birefringent medium, the two pumps could differ in polarization~\cite{Christensen_2016} rather than wavelength, thus avoiding the need for laser beams at two wavelengths. Also, the use of crystalline media where the Raman gain exhibit narrowband peaks (as opposed to silica) would enable the elimination of Raman background in the photon-pair spectrum and thus reduce noise levels significantly.

\section*{APPENDIX: Determination of the Effective Purity in the Presence of Noise}
In this Appendix we derive the inequalities in Eq.~(\ref{eq:purity_ineq}) that establish upper and lower bounds for the true quantum state purity of the signal and idler photons from measurements of the unconditional auto-correlation function $g^{(2)}_{s s^\prime}$ on the signal arm in the presence of noise.

\subsection*{Spurious noise}
We first consider the effect of noise that is generated by the interaction of either pump and creates spurious photons at the signal arm together with an additional boson -- this boson could be another photon (e.g., through single-pump SFWM), or a collective excitation in the medium (such as a phonon). We assume that before either of the pump pulses enters the fiber medium, the signal, idler, and any relevant collective excitation in the matter are in the vacuum state $\ket{\text{vac}}$. The final state after the two pump pulses leave the fiber is given by~\cite{mauerer_how_2009}
\begin{equation}
\begin{split}
\ket{\Psi}= &\Exp{\large[\beta\int d\nu_sd\nu_i f(\nu_s,\nu_i)\hat{a}^{\dagger}_{s}(\nu_s)\hat{a}^{\dagger}_{i}(\nu_i)}\\
&+\gamma\int d\nu_s d\Omega g(\nu_s,\Omega)\hat{a}^{\dagger}_{s}(\nu_s)\hat{b}^{\dagger}(\Omega)\large]\ket{\mathrm{vac}} \textrm{,}
\end{split}
\end{equation}
where $\beta$ and $\gamma$
are the amplitudes of the dual-pump SFWM and noise generation, respectively, $\hat{b}^\dagger(\Omega)$ is the creation operator of a boson with properties tagged by $\Omega$ and $g(\nu_s, \Omega)$ is the joint amplitude of the noise photon at the signal mode and the boson that is created in the interaction. We assume that all interactions are weak, i.e., $|\beta|^2$,$|\gamma|^2\ll1$. The first order in $\gamma$, $\beta$ is the lowest-order non-vacuum state, given by
\begin{equation}
\ket{\Psi^{(1)}} = \beta\ket{\psi_{si}} + \gamma\ket{\psi_{sb}} ,
\end{equation}
where
\begin{subequations}
\begin{align}
\ket{\psi_{si}} &= \int d\nu_sd\nu_i f(\nu_s,\nu_i)\hat{a}^{\dagger}_{s}(\nu_s)\hat{a}^{\dagger}_{i}(\nu_i)\ket{\text{vac}} ,\\ 
\ket{\psi_{sb}} &= \int d\nu_s d\Omega g(\nu_s,\Omega)\hat{a}^{\dagger}_{s}(\nu_s)\hat{b}^{\dagger}(\Omega)\ket{\text{vac}}, 
\end{align}
\end{subequations}
are the states associated with the creation of a signal-idler pair through the dual-pump SFWM interaction ($\ket{\psi_{si}}$) and a signal-boson pair created by spurious processes ($\ket{\psi_{sb}}$). The signal density matrices associated with these states are $\rho_s = \mathrm{Tr}_i(\ket{\psi_{si}}\bra{\psi_{si}})$ for the photon-pair state and $\rho_\text{spu} = \mathrm{Tr}_b(\ket{\psi_{sb}}\bra{\psi_{sb}})$ for the spurious photons. The auto-correlation second-order coherence can be evaluated to yield \cite{mauerer_how_2009,spring_-chip_2013}
\begin{equation}
\begin{split}
\tilde{g}^{(2)}_{ss^\prime}&=\frac{\iint d\nu_s d\nu_s^\prime\bra{\Psi}\hat{a}^{\dagger}_{s}(\nu_s)\hat{a}^{\dagger}_{s}(\nu_s^{\prime})\hat{a}_{s}(\nu_s)\hat{a}_{s}(\nu_s^{\prime})
\ket{\Psi}}{\left|\int d\nu_s \bra{\Psi}\hat{a}^{\dagger}_{s}(\nu_s)\hat{a}_{s}(\nu_s)\ket{\Psi}\right|^2}\\
&= 1 + \tilde{P}_\mathrm{raw},
\end{split}
\end{equation} 
where
\begin{equation}
 \tilde{P}_\mathrm{raw}=w^2P+(1-w)^2P_{\text{spu}}+2w(1-w)\Tr{(\rho_s\rho_{\text{spu}})}
\end{equation}
is the raw measured purity, $P = \mathrm{Tr}\rho_s^2$ and $P_\text{spu} = \mathrm{Tr}\rho_\text{spu}^2$ are the state purities of the signal and spurious photons, respectively, and $w=|\beta|^2/(|\beta|^2 + |\gamma|^2)$ is the ratio of the number of signal photons generated through dual-pump SFWM to the total number of photons. Since $0 \le \text{Tr}(\rho_s \rho_\text{spu}) \le \sqrt{P P_\text{spu}}$~\cite{MINYANG2001327}, we can derive upper and lower bounds for the measured purity:
\begin{equation}\label{eq:P_raw_tilde_ineq}
\begin{split}
\tilde{P}_{\mathrm{raw}} \ge &\; w^2 P + \left( 1-w \right)^2 P_{\text{spu}},\\
\tilde{P}_{\mathrm{raw}} \le &\; w^2 P + \left( 1-w \right)^2 P_{\text{spu}} + 2w\left( 1-w \right)\sqrt{P P_\text{spu}}.
\end{split}
\end{equation}
When $P_\text{spu}=0$ or $w=0,1$, the two bounds merge and the equality holds.

\subsection*{Detection noise}
Dark counts and ambient light that reaches the detectors constitute false detections that add background counts to the counts associated with photons that are created in the fiber. To model the effect of this type of noise we refer to the experimental setup in Fig.~\ref{fig:setup}(b), where $\textrm{APD}_s$ and $\textrm{APD}_{s^\prime}$ are used for signal auto-correlation measurements. Let us designate $p_s$, $p_{s^\prime}$ and $p_{s s^\prime}$ as the probabilities of detection events at $\text{APD}_{s}$, $\text{APD}_{s^\prime}$ and coincidences between the two, respectively, after an interaction with a single pair of dual pumps, in the absence of detection noise. Similarly, we designate $q_s$, $q_{s^\prime}$ and $q_{s s^\prime}$ as the probabilities of detecting noise events (which can be measured when both pumps are blocked) at $\text{APD}_{s}$, $\text{APD}_{s^\prime}$ or related coincidences between the two, respectively. All probabilities are assumed to be much smaller than $1$, allowing perturbative calculations. By definition, $\tilde{g}^{(2)}_{s s^\prime}=p_{s s^\prime}/p_s p_{s^\prime}=1+\tilde{P}_\mathrm{raw}$. Similarly, we define $P_\text{det}=q_{s s^\prime}/q_s q_{s^\prime} -1$. It follows then that the experimental auto-correlation, which includes photon-pair generation, spurious noise, and detection noise, is given by:
\begin{equation}\label{eq:P_raw_ineq}
\begin{split}
g^{(2)}_{s s^\prime} &= \frac{p_{s s^\prime} + q_{s s^\prime} + p_s q_{s^\prime}+ p_{s^\prime} q_s}{\left( p_s + q_s\right) \left( p_{s^\prime} + q_{s^\prime}\right)} \\
&= 1 + \left( 1-v_s \right) \left( 1-v_{s^\prime} \right) \tilde{P}_\mathrm{raw} + v_s v_{s^\prime} P_\text{det},
\end{split}
\end{equation}
where $v_{s (s^\prime)} = q_{s (s^\prime)}/(p_{s(s^\prime)} + q_{s(s^\prime)})$ is the ratio of detection noise counts to total counts on $\text{APD}_{s (s^\prime)}$. Defining $P_\mathrm{raw}=g^{(2)}_{s s^\prime}-1$ as the raw purity measured in experiment that includes noise contributions, and using Eqs.~(\ref{eq:P_raw_tilde_ineq}), Eq.~(\ref{eq:P_raw_ineq}) turns into the inequalities
\begin{equation}
\small
\begin{split}
{P}_{\mathrm{raw}} \ge &\; \left( 1-v_s \right) \left( 1-v_{s^\prime} \right) \left( w^2  P + \left(1 - w \right)^2 P_{\text{spu}} \right) + v_s v_{s^\prime} P_\text{det},\\
{P}_{\mathrm{raw}} \le &\; \left( 1-v_s \right) \left( 1-v_{s^\prime} \right) \left( w^2  P + \left(1 - w \right)^2 P_{\text{spu}} \right) + v_s v_{s^\prime} P_\text{det} \\
&+2w\left(1 - w \right) \left( 1-v_s \right) \left( 1-v_s^\prime \right) \sqrt{P P_{\text{spu}}}.
\end{split}
\end{equation}
When the two pump pulses are far delayed from each other, no dual-pump SFWM takes place, $w=0$ and the equality holds for $P_\text{raw}$; we call the value of $P_\text{raw}$ in this case the ``purity'' of the total noise, designated as $P_{\text{noise}} = (1-u_s)(1-u_{s^\prime}) P_{\text{spu}
} + u^2 P_\textrm{det}$, where $u_{s(s^\prime)}=v_{s(s^\prime)}/((1-w)(1-v_{s(s^\prime)})+v_{s(s^\prime)})$ is the ratio of detection noise counts to the total noise counts on $\text{APD}_{s(s^\prime)}$, and $u=\sqrt{u_s u_{s^\prime}}$. We further define $r_{s(s^\prime)}=w(1-v_{s(s^\prime)})$ as the fraction of dual-pump SFWM signal photon counts on $\text{APD}_{s(s^\prime)}$ to the total counts on $\text{APD}_{s(s^\prime)}$, and $t_{s(s^\prime)}=1-r_{s(s^\prime)}$ as the fraction of noise counts to the total counts on $\text{APD}_{s(s^\prime)}$. The above inequalities then become:
\begin{equation}
\begin{split}
{P}_{\mathrm{raw}} \ge &\; r^2 P + t^2 \left( 1-u_s \right) \left( 1-u_{s^\prime} \right) P_{\text{spu}} + t^2 u^2 P_\text{det},\\
{P}_{\mathrm{raw}} \le &\; r^2 P + t^2 \left( 1-u_s \right) \left( 1-u_{s^\prime} \right) P_{\text{spu}} + t^2 u^2 P_\text{det}\\
&+ 2rt\sqrt{\left( 1-u_s \right) \left( 1-u_{s^\prime} \right)} \sqrt{P P_{\text{spu}}},
\end{split}
\end{equation}
where we have defined $r = \sqrt{r_s r_{s^\prime}}$ and $t = \sqrt{t_s t_{s^\prime}}$. In terms of $P_\text{noise}$ the inequality can be rewritten as:
\begin{equation}\label{eq:P_raw}
\begin{split}
P_{\mathrm{raw}} \ge &\; r^2 P + t^2 P_\text{noise},\\
P_{\mathrm{raw}} \le &\; r^2 P + t^2 P_\text{noise} + 2 rt \sqrt{P \left(P_\text{noise} - u^2 P_{\text{det}} \right)},
\end{split}
\end{equation}
and therefore, yields the upper and lower bound of the true signal photon purity as (Eq.~(\ref{eq:purity_ineq}))
\begin{equation}\label{eq_app:purity_ineq}
\begin{split}
P \le &\;\frac{P_{\mathrm{raw}} - t^2 P_{\mathrm{noise}}}{r^2},\\
P \ge &\;\frac{P_{\mathrm{raw}} - t^2 P_{\mathrm{noise}}}{r^2} - \frac{2t}{r^2}\sqrt{P_{\mathrm{raw}}\left(P_\mathrm{noise} - u^2 P_{\mathrm{det}} \right)}.
\end{split}
\end{equation}

\section*{Funding Information}
National Science Foundation (NSF) (1521110, 1640968, 1806572).\\

%

\end{document}